# Termination control of magnetic coupling at a complex oxide interface

S. Das,[1,2,#] A. D. Rata,[1] I. V. Maznichenko,[1] S. Agrestini,[3] E. Pippel,[4] K. Chen,[5] S. M. Valvidares,[6] H. Babu Vasili,[6] J. Herrero-Martin,[6] E. Pellegrin,[6] K. Nenkov,[2] A. Herklotz,[7] A. Ernst,[4] I. Mertig,[1,4] Z. Hu,[3] K. Dörr[1,2,]*

[1] Institute of Physics, Martin Luther University Halle-Wittenberg, 06099 Halle, Germany

[2] IFW Dresden, Institute for Metallic Materials, Postfach 270116, 01171 Dresden, Germany

[3] Max Planck Institute for Chemical Physics of Solids, Nöthnitzer Strasse 40, 01187 Dresden, Germany

[4] Max Planck Institute of Microstructure Physics, Weinberg 2, 06120 Halle, Germany

[5] Institute of Physics II, University of Cologne, Zülpicher Strasse 77, 50937 Cologne, Germany

[6] ALBA Synchrotron Light Source, E-08290 Cerdanyola del Vallès, Barcelona, Spain

[7] Oak Ridge National Laboratory, Materials Science and Technology Division, 1 Bethel Valley Road, Oak Ridge, Tennessee 37831-6056, USA

*Email: kathrin.doerr@physik.uni-halle.de (corresponding author)

[#]Email: sujitdask@gmail.com

Keywords: ferromagnetic, oxide, interface, X-ray magnetic circular dichroism, magnetic coupling, exchange bias

**Atomically flat interfaces between ternary oxides have chemically different variants, depending on the terminating lattice planes of both oxides. Electronic properties change with the interface termination which affects, for instance, charge accumulation and magnetic interactions at the interface. Well-defined terminations have yet rarely been achieved for oxides of ABO$_3$ type (with metals A, B). Here, we report on a strategy of thin film growth and interface characterization applied to fabricate the La$_{0.7}$Sr$_{0.3}$MnO$_3$-SrRuO$_3$ interface with controlled termination. Ultra-strong antiferromagnetic coupling between the ferromagnets occurs at the MnO$_2$-SrO interface, but not for the other termination, in agreement with density functional calculations. X-ray magnetic circular dichroism measurements reveal coupled reversal of Mn and Ru magnetic moments at the MnO$_2$-SrO interface. Our results demonstrate termination control of magnetic coupling across a complex oxide interface and provide a pathway for theoretical and experimental explorations of novel electronic interface states with engineered magnetic properties.**

An oxide crystal of ABO$_3$ type where A and B are metal ions consists of stacked atomic planes of AO and BO$_2$. Hence, two different atomically sharp interfaces can be formed between such oxides ABO$_3$ and A*B*O$_3$, which are AO-B*O$_2$ and BO$_2$-A*O. Distinct electronic properties are expected for these interfaces. For instance, if both oxides are insulators and the AO und BO$_2$ planes are not charge-neutral, the differently terminated interfaces carry a nominal charge of opposite sign. An example is the LaAlO$_3$/SrTiO$_3$ interface containing a quasi-two-dimensional electron system with magnetic and superconducting orderings for the TiO$_2$-LaO termination[1-4] which was not detected for the SrO-AlO$_2$ termination. Yet many reported interfaces show interdiffusion that destroys a well-defined termination. Another interface with two terminations was investigated for BiFeO$_3$-La$_{0.7}$Sr$_{0.3}$MnO$_3$ by Yu et al., where the ferroelectric switching field of BiFeO$_3$ has been controlled.[5] Contrary to the experimental situation, *ab initio* theory typically assumes smooth interfaces with a well-defined termination (e. g., Ref.6). Rarely, the influence of both terminations is studied. Some recent experimental work has avoided the ambiguity by concentrating on interfaces with A*= A or B* = B.[7-9] In all other systems where A*≠ A and B* ≠ B, the interface termination is an open parameter of potentially vital impact on electronic properties. Hence, termination control of perovskite oxide interfaces is yet to be achieved and is crucial for future electronics applications of oxides. Novel electronic interface states may be discovered. Controlled magnetic interface coupling can be crucial to improve known spintronic functionalities (e. g., exchange bias). Beyond that, it has strong potential for novel device concepts, for instance, in utilizing interface control of (non-collinear) magnetic structures.

Here, we report on the $La_{0.7}Sr_{0.3}MnO_3$-$SrRuO_3$ (LSMO-SRO) interface grown with controlled terminations. Density functional theory calculations indicate that the strong antiferromagnetic Mn-Ru interface coupling of the two ferromagnetic layers is much larger for the $MnO_2$-SrO termination (called "termination 1") than for the other. X-ray magnetic circular dichroism (XMCD) measurements reveal an unusually strong magnetic coupling across this type of interface leading to rigidly coupled reversal of Mn and Ru moments at the interface. In contrast, the other interface variant ($RuO_2$-$La_{0.7}Sr_{0.3}O$, "termination 2") shows a more conventional, exchange-bias-like magnetic coupling. Our results demonstrate efficient termination control of magnetic order and coupling across a coherent oxide interface.

Growing chemically stable layers of half unit cells (of AO or $BO_2$) would be a straightforward strategy to control the interface termination. It was originally applied to $LaAlO_3/SrTiO_3$,[1] but this strategy is rarely applicable because of chemical or structural instabilities.[5] Moreover, very low interdiffusion is crucial for a well-defined interface termination. These are severe challenges for the realization of one or the other termination for many oxide interfaces, in particular for polar insulators.[4] Chemistry and growth conditions are decisive and have to be specifically explored for any combination of two oxides at the present stage. Here, layer thicknesses of 9 unit cells of LSMO and 14 unit cells of SRO were grown coherently by RHEED-controlled pulsed laser deposition (PLD) on $TiO_2$-terminated STO(001) substrates with an in-plane lattice parameter of 3.905 Å. We investigated the interfaces using High-Resolution Scanning Transmission Electron Microscopy (STEM) to analyze interdiffusion and termination (see below). An atomically sharp $(La,Sr)O-MnO_2/SrO-RuO_2$ interface (termination 1) has been found after growing LSMO and, subsequently, SRO on a $TiO_2$-terminated substrate of $SrTiO_3(001)$ (STO). If full unit cells grow, the reversed layer sequence (LSMO/SRO/STO(001)) leads to termination 2 with the $SrO-RuO_2/(La,Sr)O-MnO_2$ interface. The latter interface shows modest interdiffusion which is, however, weak enough to keep the well-defined character of the termination. We note that similar bilayers of $LaMnO_3/LaNiO_3$ with reversed layer sequence have been studied by Gibert el al.,[9] also revealing a sharper and a more interdiffused interface. The level of interdiffusion is, however, much higher in that case, leading to a novel intermixed phase at the interface. (For further structural characterization, see the Supplementary Information.)

The microstructure and, in particular, the type of termination have been examined by STEM using High Angle Annular Dark Field (HAADF) contrast (Fig.1). In this mode, the intensity at an atomic column grows approximately with the square of the ordering number Z of the atom in the periodic table. This also holds for the weighted average of Z values when a lattice site is occupied by different atoms. Interdiffusion occurs only within the same sublattice in the present system, i. e. La and Sr stay at the A sites and Mn and Ru remain at the B sites of the $ABO_3$ structure. The Z values of A site atoms are 38 (Sr) and 51 ($La_{0.7}Sr_{0.3}$), those of B site atoms are 25 (Mn) and 44 (Ru). Hence, the elemental Z contrast for the LSMO-SRO interface is strong for both, A and B sites. This favourable situation arises from the presence of a 3d (Mn) and a 4d (Ru) element on B sites. Thickness variations of STEM samples additionally change the intensity ratio for any two elements.[10] The bulk intensity values of Mn, Ru, Sr and $La_{0.7}Sr_{0.3}$ columns have been estimated by averaging intensities at atomic positions away from the interface. For termination 1, the intensity profile along a line crossing the interface and all types of atomic columns clearly shows the $MnO_2$-SrO interface (Fig.1a). Mn and Sr intensities at the interface are in agreement with the bulk intensity values within the error range (~10%). This result indicates that this interface termination is chemically stable. For termination 2, the $RuO_2$-(La,Sr)O interface is clearly detectable, even though it shows weak interdiffusion (Fig.1b). The Mn peak next to the interface is larger than in the bulk, while the Ru site at the interface is slightly reduced in intensity, both indicating some Mn/Ru interdiffusion. Intermixing at the A sites is not detected. Averaging over 10 parallel lines crossing the interfaces in different places confirm the results shown in Fig.1. We note that the $RuO_2$ top layer of the SRO film for termination 2 is not in agreement with expectations, because a SrO termination has been reported for the surface of a single SRO film.[11] In general, the chemical stability of an interface may differ from that of a free surface. (More discussion of the present case in the Supplement.)

An *ab initio* electronic structure calculation based on density-functional theory (DFT) has been employed to study the two LSMO-SRO interface terminations. A Green's-function method within a multiple-scattering expansion has been used.[12,13] We consider superlattices consisting of 8 unit cells (uc) of LSMO, 8 uc of SRO, and 4 uc of STO. The superlattices are constructed either with SrO-$MnO_2$ (termination 1) or with $RuO_2$-($La_{0.7}Sr_{0.3}$)O (termination 2) at the LSMO-SRO interface, and all layers consist of complete unit cells (Fig.1, bottom panel).

In order to verify that using a LSMO-SRO-STO superlattice in the calculation is appropriate, the effect of an STO cap layer on the bilayers has experimentally been checked and was found to have negligible influence on the magnetic behaviour (Supplementary Information). The groundstate of both, the SRO and the LSMO layer exhibits ferromagnetic ordering. This includes robust ferromagnetic order also of the last $MnO_2$ layer at the LSMO-SRO interface for both terminations. Across the interface, the SRO and LSMO layers are coupled antiferromagnetically (AFM) for both terminations. Notably, the calculated coupling constant *J* between Mn and Ru across the interface is larger by about a factor of 3 for termination 1 (Tab. 1). Furthermore, the difference of the total energy ($\Delta E$) between the states with either antiferromagnetic or ferromagnetic coupling across the interface was determined and confirms the much higher stability of the antiferromagnetic state for termination 1 (Tab. 1). These results suggest that the antiferromagnetic coupling is much stronger for the interface termination 1.

The magnetic interface coupling has been investigated by measuring X-ray magnetic circular dichroism (XMCD) spectra of the Mn $L_{2,3}$ and Ru $L_3$ edges. Mn- and Ru-XMCD $L_3$ edge spectra have been recorded as a function of an external magnetic field at 60 K with the monochromatic photon beam at grazing (*H //* [100], in-plane) and normal (*H //* [001], out-of-plane) incidence yielding element-specific magnetic hysteresis loops (Fig.2; out-of-plane data in the Supplement, Figs.S3-S5). This data provides insight into the individual switching processes of Mn and Ru magnetic moments. In the following, hysteresis loops are discussed based on Fig.2 showing field-dependent in-plane Mn- and Ru-XMCD intensities, starting with the more conventional case of termination 2. The in-plane Ru hysteresis is that of a typical ferromagnet (Fig.2c). The respective out-of-plane hysteresis loop (Fig.S4a) has substantial remanence, too, indicating a canted orientation of Ru magnetic moments as sketched in the layer schemes in the figures. This behavior is consistent with previous magnetization[14] and XMCD[15] results from orthorhombic SRO/STO(001) films. In-plane Mn-XMCD hysteresis data (Fig.2e) show an inverted loop with negative (positive) remanence after applying a large positive (negative) field. This reflects the antiferromagnetic coupling of Mn to the Ru moments which are aligned in field direction (Fig.2c) and cause the Mn moments to reverse in a positive magnetic field. The behavior can be described by a positive exchange bias effect.[16,17] The out-of-plane Mn hysteresis (Supplement, Fig.S4b) has significant remanence, too, and in-plane Mn data (Fig.2e) show a gradual intensity decrease in a reducing in-plane field below 1 T. Both findings reflect an out-of-plane canting of Mn moments induced by the antiferromagnetic interface coupling. (A single LSMO/STO(001) film has magnetic in-plane easy axes.[18]) In strong contrast to this exchange-bias-like characteristics, the termination 1 sample shows an abrupt reversal of Mn moments in the film plane ($H//$[100]) (Fig.2d), with a small saturation field of about 30 mT. The respective out-of-plane Mn hysteresis loop has negligible remanence (Fig.S5b). Hence, the measured [100] direction is a magnetic easy axis of LSMO, and no out-of-plane canting of Mn is present. This is in conflict with significant magnetic coupling of Mn moments to fixed Ru moments at the interface, because the Mn switching field would be enlarged by such coupling. However, the associated in-plane Ru-XMCD hysteresis loop (Fig.2b) has a peculiar shape: a gradual intensity decrease appears when the positive magnetic field is reduced to zero, followed by a striking *increase* when the magnetic field turns negative. We argue that these features indicate the presence of an interface-near layer of Ru moments that switches in rigidly coupled way with the Mn moments. The maximum field of 3.5 T at 60 K (Fig.2b) is sufficient to break the antiferromagnetic coupling, aligning Ru and Mn moments along the field direction. The decreasing Ru intensity with reducing positive field reflects an increasing antiparallel alignment of Ru to Mn moments at the interface. The enhancement of the Ru intensity in small negative field originates from the coupled reversal of antiferromagnetically aligned Ru and Mn moments at the interface, pointing these Ru moments in the positive direction again. On the other hand, there is an upper part of the SRO layer which remains oriented to the external field during this process, reflected by the fact that the total Ru intensity stays positive at $H = 0$ (Fig.2b). The Ru in this upper part of the SRO layer is canted out-of-plane as is detectable in the substantial remanence of the out-of-plane Ru-XMCD hysteresis loop (Fig.S5a), and, thus, behaves similar to the SRO layer of the termination 2 sample. The Ru canting is not extended to the interface for termination 1, because the coupled Mn shows no canting, in contrast to the earlier discussed case of termination 2. These observations lead to two crucial conclusions: (i) Near the interface of termination 1, antiferromagnetically coupled Mn and Ru moments are aligned in-plane along the directions of [100] (and [010], for symmetry reasons) and switch in coupled way in a magnetic field. This is not the case for the interface of termination 2 where an exchange bias effect of Ru on Mn moments is observed. (ii) For the SRO layer in the termination 1 sample, a depth-dependent rotation of Ru moments is concluded in similarity to the exchange-spring effect,[19-21] in agreement with a recent suggestion in a polarized-neutron study of Kim et al.[22] It seems likely that the upper part of the SRO layer returns to the magnetic order and orthorhombic lattice symmetry of a single SRO/STO(001) film,[14,23] since the influence of the interface decays with growing distance. At the interface, SRO has in-plane [100] and [010] magnetic easy axes in agreement with the magnetic anisotropy observed in tetragonal SRO films under larger in-plane strain (i. e., at an in-plane lattice parameter $a \geq 3.93$ Å [24]), suggesting a symmetry change of the interface-near SRO atomic layers. The strong magnetic exchange coupling at the interface of termination 1 seems to induce a non-collinear spin structure in the SRO layer which can be controlled and switched by the adjacent magnetic layer (LSMO) using moderate magnetic fields.

More insights into the effect of interface termination on magnetic ordering have been obtained from magnetization (*M*) measurements (Fig.3 and Supplement). LSMO is a soft ferromagnet which orders at $T_c^{LSMO}$ = 370 K. Coherently grown LSMO films on STO(001) experience biaxial tensile strain that reduces $T_c^{LSMO}$.[25,26] The few unit cells thick LSMO layer is subject to further reduction of $T_c^{LSMO}$ due to the finite size and interface effects.[18] In our samples, the 9 uc thick LSMO layer orders at $T_c^{LSMO} \sim$ 310 K, and $T_c^{LSMO}$ is systematically larger by ~10 K in samples with termination 1. $T_c^{SRO} \sim$ 140 K is the Curie temperature of the SRO layer for both termination types. SRO grown coherently on STO(001) is orthorhombic with a canted, temperature-dependent out-of-plane easy axis.[14] Magnetic ordering of LSMO-SRO bilayers or superlattices is complex at $T < T_c^{SRO}$,[16,17,22,27-30] since four energy scales may be dominant depending on interface quality, layer thickness, temperature, magnetic field and magnetic history. These are the antiferromagnetic coupling at the interface, magnetic anisotropy energy, magnetostatic (stray field) energy and Zeeman energy in an external magnetic field. For Fig.3, samples have been measured during cooling to 20 K in various in-plane magnetic fields $H_{FC}$ // [100] and subsequent warming in a small field (0.1 T). Cooling in 0.1 T leads to a distinct drop of *M* at $T_c^{SRO}$ when Ru moments start to align antiparallel to Mn moments. Increasing the cooling field provides a measure for the coupling strength, because the drop of *M* at $T_c^{SRO}$ changes into a rise at a threshold field. A cooling field of 1 T suppresses the *M* drop at $T_c^{SRO}$ in case of termination 2 (Fig.3b). For termination 1, the slope (d*M*/d*T*) change at $T_c^{SRO}$ indicates antiferromagnetic alignment of some Ru moments in a field of 1 T, confirming the larger coupling strength. (Note also the difference of warming curves for 1 and 3 T (Fig.3a), whereas there is no difference for the respective curves for termination 2 (Fig.3b).) The total magnetization recorded during cooling grows with $H_{FC}$, indicating a gradually increasing alignment of both, Mn and Ru moments with the field against the antiferromagnetic interface coupling. When the field is reduced to 0.1 T at 20 K, the magnetization drops as expected since Ru and Mn moments return to antiferromagnetic alignment at the interface. Interestingly, the observed reduction of $M$ grows with $H_{FC}$ (Fig.3a,b). We note that from the magnetization data it is not possible to decide whether Mn or Ru moments reverse upon removal of the cooling field. Ru moments may be fixed in the crystal lattice because of the large magnetic anisotropy of SRO at 20 K.[14,24] Consequently, Mn would reverse in relation to the amount of Ru moments at the interface which aligned to the field during cooling. Data in Fig.2b, taken at 60 K, reveal an alternative mechanism: the Ru interface layer reverses instead of the LSMO layer. A qualitatively similar behavior of cooling and warming curves is observed for interface termination 2 (Fig.3b) and can be understood in similar way based on the antiferromagnetic interface coupling. Here, one can conclude Mn reversal to occur at 20 K upon switching off the cooling field, because this agrees with the Mn-XMCD hysteresis loop at 60 K (Fig.2e), and the lower temperature enhances the magnetocrystalline anisotropy of SRO. Besides this, canting of Mn and Ru moments causes additional magnetization rotation into the film plane for termination 2.

Another obvious difference of termination 1 and 2 samples which is, however, not arising from the LSMO-SRO interface is the value of the saturated moment ($M_S$) of LSMO (Fig.3). Magnetization loops (Fig.3a and Fig.3b, insets) recorded along the [100] in-plane direction at 170 K (i. e., well above $T_C^{SRO}$) have been used to estimate $M_S$ of LSMO. (Since the high-field slope of $M(H)$ is subject to error from the uncertainty of the substrate contribution (see Methods part), the magnetization value at 0.2 T has been taken for estimating $M_S$.) Termination 1 shows $M_S$ ~1.0 $\mu_B$/uc (2.6 $\mu_B$/Mn) at 170 K, compared to 0.6 $\mu_B$/uc (1.7 $\mu_B$/Mn) for termination 2. (All magnetization values have an error of < 10 %.) The difference of ~0.4 $\mu_B$/uc between termination 1 and 2 samples is systematically observed in several sample pairs of both terminations. We attribute it to higher Mn order in termination 1 samples, which is not located at the LSMO-SRO interface, but at the other side of the LSMO layer. (Besides the theoretical result of robust ferromagnetic Mn order at the LSMO-SRO interface, numerous investigations on superlattices (e. g., Ref.30) have shown that an interface to SRO stabilizes ferromagnetic order in LSMO.) Antiferromagnetic alignment of one atomic Mn layer to the remainder of the LSMO film results in a change of $M_S$ of 0.3 $\mu_B$/uc (or 0.7 $\mu_B$/Mn, averaged over the 9 uc of LSMO). This approximately agrees with the observed difference of ~0.4 $\mu_B$/uc. X-ray absorption spectroscopy of LSMO films with free surface or capped with STO pointed to the presence of an antiferromagnetic Mn layer,[31] in agreement with our result for termination 2 where the LSMO layer has a free surface. The large $M_S$ of termination 1 samples, on the other hand, implies the absence of an antiferromagnetic Mn layer at the substrate interface of LSMO. In the context of recent work on the LSMO/STO(001) interface,[32,33] this indicates a subtle balance of the Mn order which can be ferro- or antiferromagnetic depending on precise chemical composition at the interface and, possibly, further parameters.

Our experimental and theoretical results are in excellent agreement and provide a strong foundation for understanding the achieved realization of termination controlled magnetic coupling. For completion of the discussion, we briefly consider the effect of elastic coupling through rotations of oxygen octahedra across the interface.[23,34,35] Linked with that is the question about the impact of the growth sequence of the bilayer samples. Octahedral rotations are governed by the elastic strain of a film and, additionally, can be induced at a coherent interface between two oxides with a penetration depth of few unit cells.[23,34,36,37]

SRO as a single coherently grown film on STO(001), i. e., in the same epitaxial strain state as in the present samples, is tetragonal at the growth temperature of 700°C.[38] It undergoes a phase transition to orthorhombic *Pbnm* symmetry (with the Glazer notation $a^-a^-c^+$ of octahedral rotations) at about 280°C. The different symmetries are associated with distinct rotational patterns of oxygen octahedra in SRO.[23,39] LSMO on STO(001) shows the $a^+a^-c^0$ rotational pattern.[23,36] Thus, one expects an influence of the transfer of octahedral rotations between LSMO and SRO on the interfacial lattice structure and, consequently, the magnetic ordering and coupling. However, since the SRO layer undergoes a structural transition during cooling far below the growth temperature, the elastic interaction between LSMO and SRO is independent of the growth sequence during the transition. This holds under the assumption that the transfer of octahedral rotations induced by the substrate is negligible at the position of the interface. The latter is fulfilled because of the layer thicknesses of 9 or 13 uc, respectively, of the first layer on the substrate. (For further discussion, see the Supplementary Information.) Hence, the differences observed for the two terminations do not originate from the elastic transfer of octahedral rotations. On the other hand, a different oxygen rotational pattern is likely to occur for the two terminations of the LSMO-SRO interface, considering the differences of magnetic anisotropy of Mn and Ru moments at both interface types.

In summary, magnetic ordering and coupling across the LSMO-SRO interface is strongly different for the two interface terminations as is found by combining experimental and theoretical efforts. The $MnO_2$-SrO termination shows unusually strong antiferromagnetic Mn-Ru coupling in such manner that magnetic switching proceeds by rigidly coupled reversal of Mn and Ru moments at the interface. This phenomenon leads to a non-collinear spin structure inside the SRO layer which can be controlled by a moderate magnetic field. These findings have strong implications for spintronics devices utilizing the exchange-bias effect in magnetic oxide layers. Beyond that, the control of non-collinear spin textures at interfaces may provide a valuable tool for cutting-edge spintronics approaches where the motion of domain walls or other chiral spin textures is exploited for a new generation of fast electronic devices.[40,41] Density functional theory proved to be an efficient tool to reveal the impact of termination on electronic properties of oxide interfaces and may be applied to predict other (ABO$_3$) oxide interfaces with strong influence of termination on magnetic order or, more generally, the nature of the electronic state at the interface. From experimental point of view, the stability of the termination variants is unknown for most pairs of ABO$_3$ oxides and might be explored by single-interface growth and STEM-based advanced detection tools including atomically resolved Electron Energy Loss Spectroscopy (EELS). Termination control in thin film heterostructures of complex oxides clearly promises novel electronic interfaces states and, thus, holds a strong potential for a new generation of oxide electronics.

**Methods**

***Thin film growth.*** Bilayers with layer thicknesses of 9 unit cells of LSMO and 14 unit cells of SRO were grown by RHEED-controlled pulsed laser deposition (PLD with laser wavelength of 248 nm) from stoichiometric targets. The single-crystalline SrTiO$_3$ (STO) substrates were (001) oriented and TiO$_2$ – terminated. The growth temperature and laser energy were 700°C and 0.3 J / cm$^2$, respectively, and the oxygen pressure in the chamber was 0.2 mbar. After growth, samples were annealed for 1 hour in oxygen of 200 mbar. Structural characterization is reported in the Supplementary Information.

***Scanning Transmission Electron Microscopy.*** All STEM measurements were performed at a Cs probe-corrected FEI TITAN operating at 300 kV equipped with a Fischione HAADF detector. The cross-sectional samples were prepared using standard mechanical polishing and dimpling techniques with a final polishing in a Gatan PIPS ion mill using a 3 kV argon beam.

***Density functional theory.*** First-principles calculations were performed using a self-consistent Green´s function method especially designed for semi-infinite systems such as surfaces and interfaces.[12,13] A generalized gradient approximation (GGA) was utilized to describe correctly electronic properties of La$_{0.7}$Sr$_{0.3}$MnO$_3$-SrRuO$_3$ interfaces. This functional reproduces adequately electronic and magnetic properties of SrRuO$_3$.[42] 3d states of Mn in La$_{0.7}$Sr$_{0.3}$MnO$_3$ were additionally corrected using a GGA+U approximation.[43] The effective U was chosen to be 0.9 eV, which provides the correct magnetic moment of Mn and magneto-optical spectra of La$_{0.7}$Sr$_{0.3}$MnO$_3$.[44] The alloying of Sr and La in La$_{0.7}$Sr$_{0.3}$MnO$_3$ was simulated with a coherent potential approximation as it is formulated within the multiple scattering theory.[45]

***X-ray Magnetic Circular Dichroism.*** The XMCD experiments were performed at the BL29 BOREAS beamline at the ALBA synchrotron radiation facility in Barcelona. The XAS was measured using circular polarized light with the photon spin parallel ($\sigma^+$) or antiparallel ($\sigma^-$) with respect to the magnetic field. The spectra were collected with the beam in grazing (*H* // film surface) and in normal (*H* ⊥ film surface) incidence.[15] The degree of circular polarization delivered by the Apple II-type elliptical undulator was 70% (close to 100%) for the Ru L-edge (Mn L-edge). The spectra were recorded using the total electron yield method (by measuring the sample drain current) in a chamber with a vacuum base pressure of $2 \times 10^{-10}$ mbar. The XMCD data were measured at *T* = 60 K after cooling the samples in a field of 3 Tesla. The Ru-XMCD hysteresis loops were obtained by measuring the Ru $L_3$ edge XMCD spectrum for each value of magnetic field and integrating the XMCD signal. The Mn-XMCD hysteresis loops were obtained by measuring, as a function of applied field, the Mn $L_3$ edge XMCD signal at the energy where the XMCD signal is maximum.

**Magnetization measurements**. Magnetization at temperatures between 20 K and 360 K has been measured in a Superconducting Quantum Interference Device (SQUID) *MPMS* Magnetometer (*Quantum Design*). The substrate contribution has been eliminated by subtracting the diamagnetic part of the magnetization derived from the negative high-field slope d*M*/d*H* of the data.

**Acknowledgements**

S. D., A. D. R., K. D., I. M., A. E. and I. V. M. acknowledge support from Deutsche Forschungs-gemeinschaft under the grant of SFB 762 *Functional Oxide Interfaces*. Discussions with M. Trassin, M. Ziese, H. M. Christen, E.-J. Guo, F. Grcondciel, M. Bibes and H. N. Lee are gratefully acknowledged.

**Contributions**

S. D. grew the films. S. D. and A. D. R. did x-ray and magnetization measurements. E. Pippel conducted the HAADF-STEM imaging and analysed the atomic configuration near the interfaces. I. V. M., I. M., and A. E. did the DFT calculations. S. A., Z. H., A. D. R. and K. C. performed the XMCD measurements. H. B. V., S. M. V, J. H. M. provided the support at the beamline. S. A. carried out the analysis of the XMCD data. A. H. helped with film growth and sample characterization. K. N. assisted in magnetization measurements. S. D., A. D. R., I. M., I. V. M., S. M. V., and E. Pellegrin discussed and K. D. wrote the manuscript.

The authors have no competing financial interests to report.

| Termination | $\Delta E$ (meV) | $J$ (meV) |
|---|---|---|
| 1 | -71.5 | -13.3 |
| 2 | -15.1 | -4.5 |

**Table 1: Antiferromagnetic coupling between $La_{0.7}Sr_{0.3}MnO_3$ and $SrRuO_3$ evaluated by density functional theory.** Difference $\Delta E = E_{AFM} - E_{FM}$ of the total energy for ferromagnetic ($E_{FM}$) or antiferromagnetic ($E_{AFM}$) coupling across the interface, and next-neighbour exchange coupling constant $J$ between Mn and Ru.

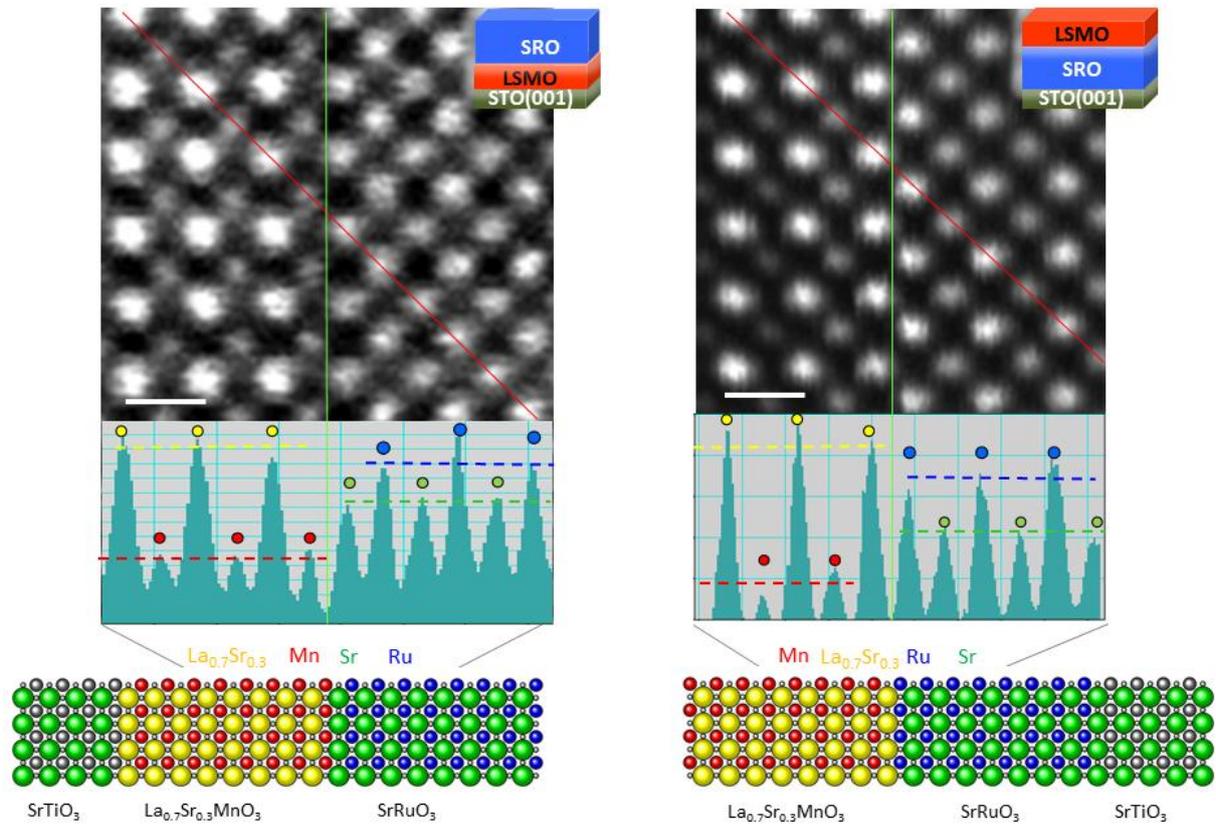

**Fig.1**. **Atomic structure of termination 1 (left) and termination 2 (right) of the La$_{0.7}$Sr$_{0.3}$MnO$_3$-SrRuO$_3$ interface.** HAADF-STEM Z-contrast images are rotated by 90° with respect to the related layer systems (upper insets). The interface is marked with a green line. The lower insets show the intensity profile recorded along the red line. Colored dashed lines indicate an averaged value away from the interface. Bottom panel: lattice structure used for DFT calculations. Scale bar length of 4 Å.

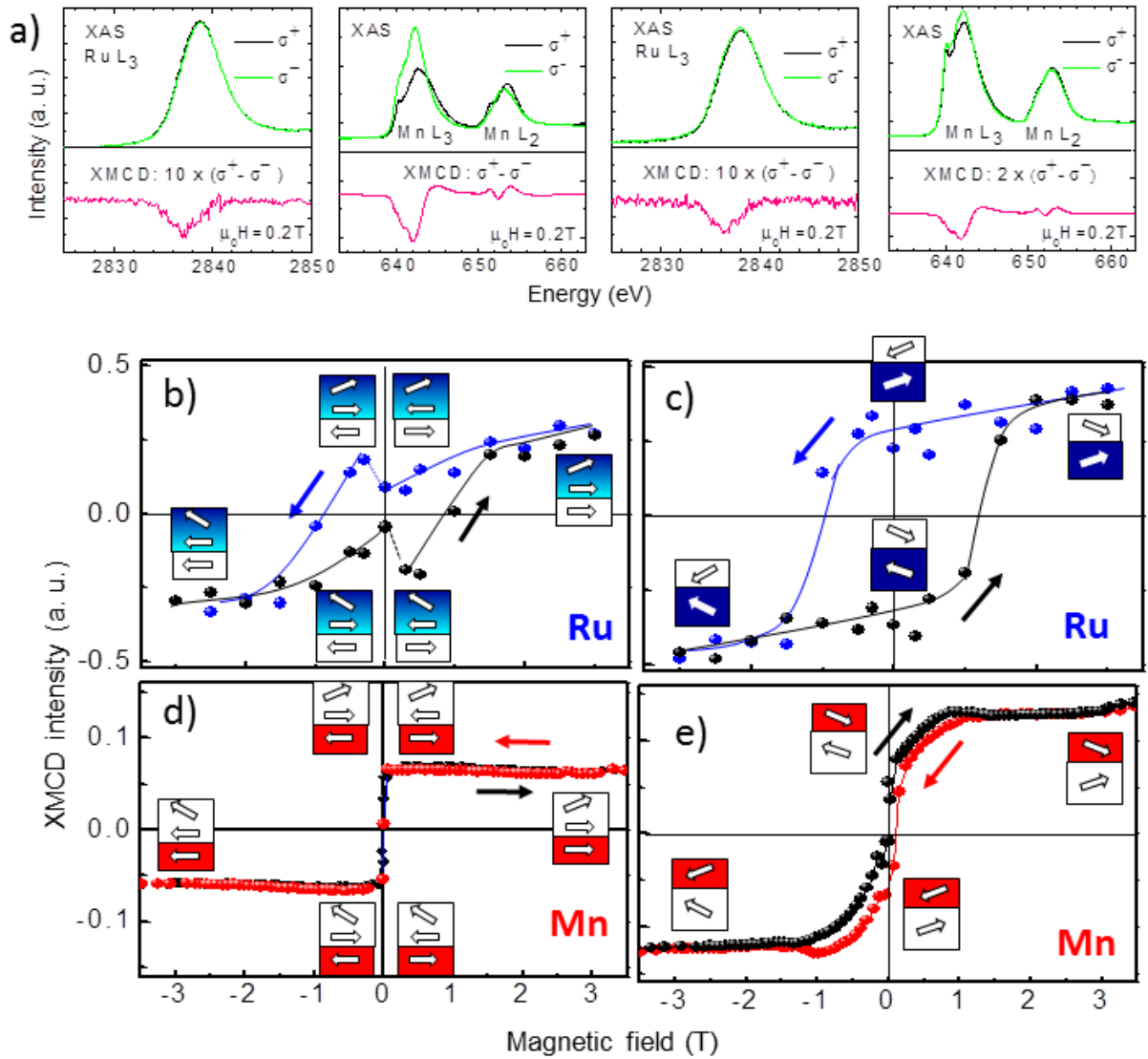

**Fig.2. Magnetic order and element-specific hysteresis loops derived from X-ray magnetic circular dichroism (XMCD) for termination 1 (left) and termination 2 (right) at 60 K.** a) Representative Ru and Mn spectra at 0.2 T recorded along the in-plane [100] direction (grazing incidence). Ru-XMCD in-plane hysteresis loop for terminations 1 (b) and 2 (c), respectively. Insets show the layer scheme with directions of Mn (red) and Ru (blue) magnetic moments. Out-of-plane canting has been derived from similar measurements along the [001] direction, see Supplementary Information. d, e) Mn-XMCD in-plane hysteresis loops in analogy to b, c).

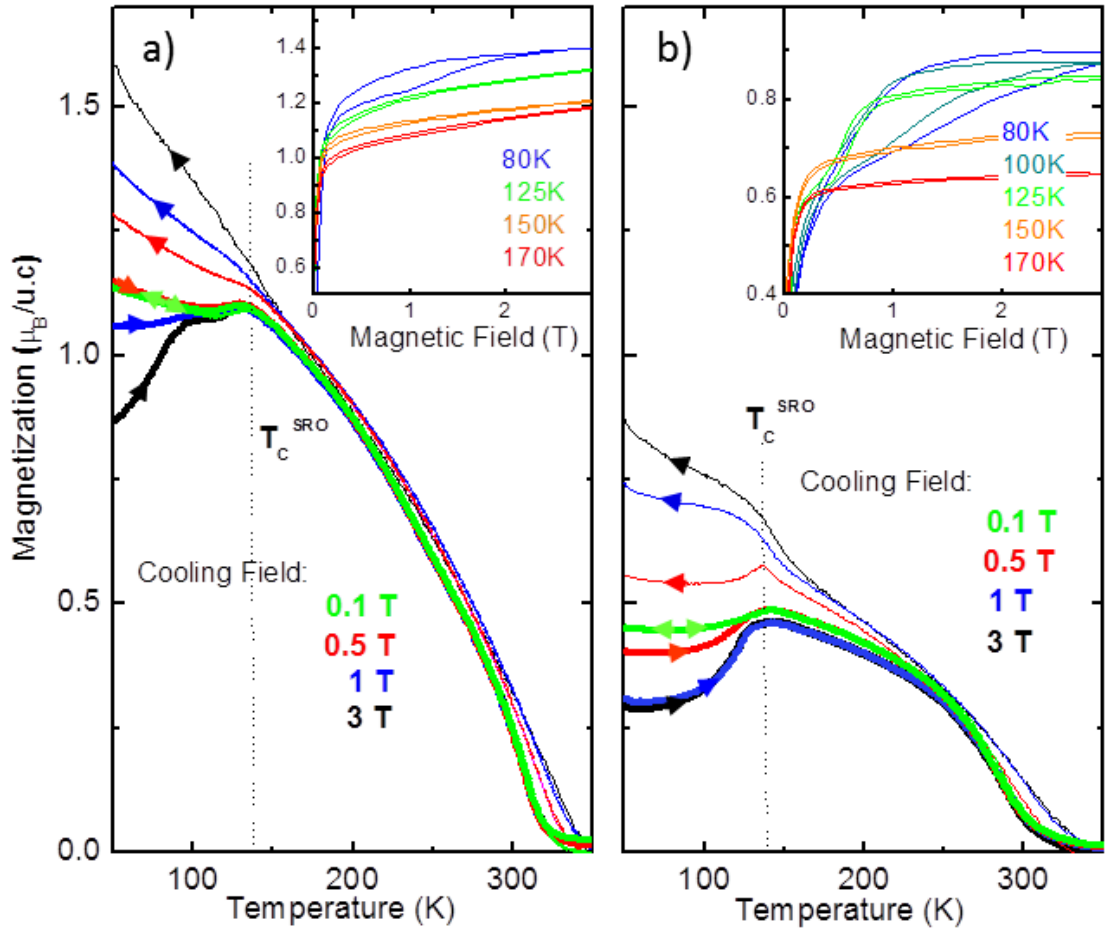

**Fig.3. Influence of magnetic field during cooling on magnetization.** Magnetization recorded in in-plane magnetic field $H$ //[100] during field cooling (thin lines, field values as indicated) and during warming in 0.1 T (thick lines) for terminations 1 (a) and 2 (b). Note the coincidence of thin and thick green lines. $T_C^{SRO}$ marks the Curie temperature of the SRO layer. Insets: Field-dependent magnetization at constant temperatures as indicated.

**Supplementary Information**

**Termination control of magnetic coupling at a complex oxide interface**

S. Das,[1,2,#] A. D. Rata,[1] I. V. Maznichenko,[1] S. Agrestini,[3] E. Pippel,[4] K. Chen,[5] S. M. Valvidares,[6] H. Babu Vasili,[6] J. Herrero-Martin,[6] E. Pellegrin,[6] K. Nenkov,[2] A. Herklotz,[7] A. Ernst,[4] I. Mertig,[1,4] Z. Hu,[3] K. Dörr[1,2,]*

[1] Institute of Physics, Martin Luther University Halle-Wittenberg, 06099 Halle, Germany

[2] IFW Dresden, Institute for Metallic Materials, Postfach 270116, 01171 Dresden, Germany

[3] Max Planck Institute for Chemical Physics of Solids, Nöthnitzer Strasse 40, 01187 Dresden, Germany

[4] Max Planck Institute of Microstructure Physics, Weinberg 2, 06120 Halle, Germany

[5] Institute of Physics II, University of Cologne, Zülpicher Strasse 77, 50937 Cologne, Germany

[6] ALBA Synchrotron Light Source, E-08290 Cerdanyola del Vallès, Barcelona, Spain

[7] Oak Ridge National Laboratory, Materials Science and Technology Division, 1 Bethel Valley Road, Oak Ridge, Tennessee 37831-6056, USA

*Email: kathrin.doerr@physik.uni-halle.de (corresponding author)

[#]Email: sujitdask@gmail.com

## 1. Thin film growth and structural characterization

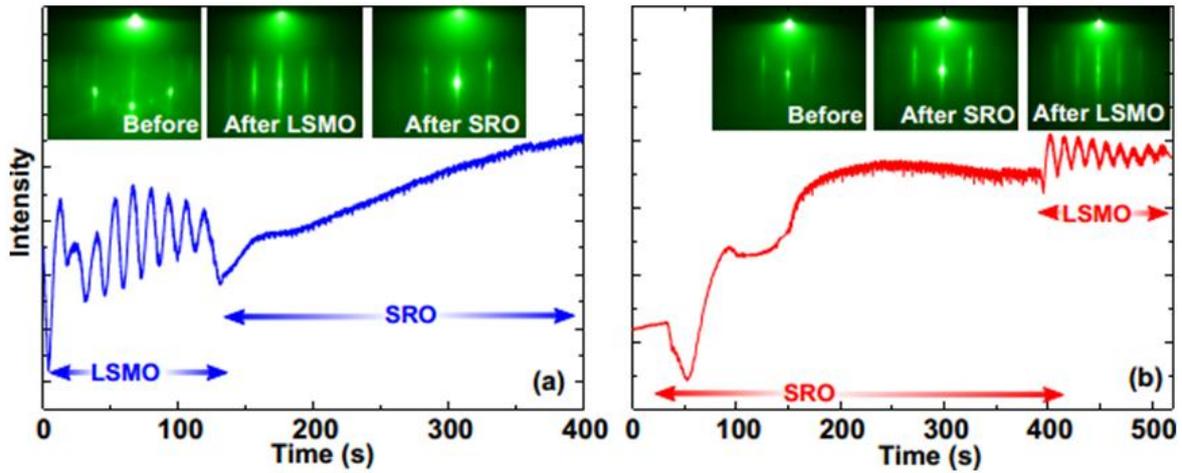

**Fig.S1: Reflection High Energy Electron Diffraction (RHEED) intensity during growth.** Time-dependent RHEED intensity for a) SRO/LSMO/STO(001) (termination 1) and b) LSMO/SRO/STO(001) (termination 2). Insets: RHEED patterns observed before growing the first layer and after finishing each layer.

$La_{0.7}Sr_{0.3}MnO_3$ (LSMO) and $SrRuO_3$ (SRO) layers have been grown by Pulsed Laser Deposition (PLD) using an excimer laser with 248 nm wavelength and a PLD chamber (*Surface GmbH*) with High-pressure Reflection High Energy Electron Diffraction (RHEED) facility. $TiO_2$-terminated $SrTiO_3$(001) substrates have been prepared as reported by Connell et al.[1] Fig.S1a, b show an example of the time-dependent RHEED intensity observed during growing both types of bilayers. LSMO grows in layer-by-layer growth mode reflected in the RHEED intensity oscillations. In termination 1 samples, the SRO layer grows on the LSMO layer without RHEED oscillations, but a smooth plateau of gradually increasing intensity is observed after the initial increase (Fig.S1a). The RHEED patterns (insets of Fig.S1a) and topography images taken by Atomic Force Microscopy (AFM) show low roughness with steps of unit cell height indicating the step-flow growth mode of SRO on LSMO. In termination 2 samples, SRO grows in the step-flow growth mode after an initial layer-by-layer stage of growth as reported earlier.[2,3] The initial RHEED intensity evolution for SRO growth on STO(001) shows two oscillations before reaching a plateau of reflectivity, in close resemblance to the result of Choi et al. (Fig.1 in Ref.2). Choi et al.[2] and Rijnders et al.[3] attribute the initial growth stage to reduced mobility of deposited ions on the substrate surface leading to initial layer-wise growth, whereas the mobility rises after completion of the first SRO layer, changing the growth mode to step-flow. Additionally, a change to SrO surface termination of the substrate has been concluded from RHEED and AFM measurements in the initial growth stage by Rijnders et al.[3] Assuming growth of full unit cells of SRO after this termination change, the SRO surface naturally ends with a SrO layer. If this is correct for our growth process irrespective of differences in the growth parameters, the question arises why it was possible to prepare the interface of termination 2 samples with RuO$_2$ termination of the SRO layer. The answer might be related to chemical or kinetic interface stability. One option is the presence of a chemical driving force that changes the SrO uppermost layer to a La$_{0.7}$Sr$_{0.3}$O layer during initial LSMO growth, moving excess Sr ions upwards into the growing LSMO film. This hypothetical mechanism does not necessarily mean there is a Sr excess in the completed film. Significant Sr excess in the LSMO layer would be reflected in a suppression of the Curie temperature which remains high (~300 K) in our samples for a 35 Å (9 uc) thick LSMO layer, excluding significant Sr overdoping. However, Sr enrichment at the top of the LSMO layer may have occurred as a termination change in similarity to the case of single SRO films,[3] i. e. the LSMO layer would not terminate with a MnO$_2$ layer at the surface, but with a SrO layer. Alternatively, the topmost (La,Sr)O layer may have strong excess of Sr. These are hypothetical scenarios requiring further clarification.

Coherent growth of both layers, LSMO and SRO, on STO(001) with in-plane lattice parameter of $a$ = 3.905 Å has been confirmed by X-ray diffraction and High-resolution Scanning Transmission Electron Microscopy (STEM). No dislocations or other lattice defects breaking lattice coherency were found by STEM imaging. X-ray reciprocal space maps recorded around the (103) reflection show relatively weak film reflections because of the low layer thicknesses. No indication of strain relaxation is detectable in the film reflections. Nominal misfits of pseudocubic lattice parameters of the layers on STO(001) are ($a_{LSMO} - a_{STO}$)/$a_{LSMO}$ = -0.75 % for LSMO (with $a_{LSMO}$ = 3.876 Å) and +0.38 % for SRO (with $a_{SRO}$ = 3.92 Å). Hence, strain relaxation is not likely for the thin layers of 9 uc LSMO and 13 uc SRO.

## 2. DFT calculations

In addition to the magnetic coupling constants discussed in the main text, the magnetic moments in the LSMO/SRO/STO heterostructures were calculated from first principles for terminations 1 and 2, respectively (Fig.S2). The local magnetic moments of Ru and Mn are shown in the lower part of Fig.S2. The dashed lines illustrate the magnetic moments of Ru and Mn in SRO and LSMO bulk, respectively. The Mn magnetic moment is slightly reduced at the SrO/MnO$_2$ interface (termination 1), whereas it keeps the bulk value at the RuO$_2$/La$_{2/3}$Sr$_{1/3}$O interface (termination 2). A plausible reason for this result is that at the SrO/MnO$_2$ interface the effective concentration of La is reduced: if one considers only the atomic planes next to the interface MnO$_2$ plane, the average composition would be La$_{0.33}$Sr$_{0.67}$ leading to a nominal oxidation state of Mn$^{3.67+}$. (This consideration is for plausibility only, because LSMO is a metal and valence states cannot strictly be derived by considering next neighbors.) At the RuO$_2$/La$_{0.67}$Sr$_{0.33}$O interface (termination 2), Mn has the same neighbors as in LSMO bulk; consistently, its magnetic moment remains rather unchanged. At this interface, Ru shows a slightly reduced magnetic moment (Fig.S2).

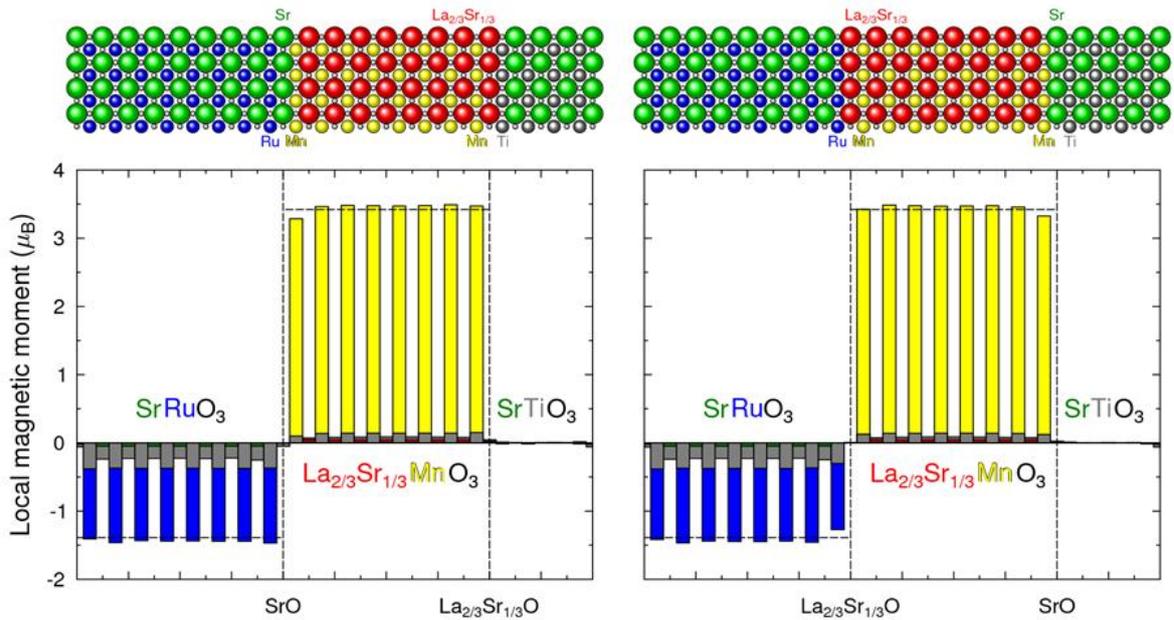

**Fig.S2. Local magnetic moments of Ru and Mn for terminations 1 (left) and 2 (right) calculated using density functional theory.** The units of the $SrRuO_3$-$La_{0.67}Sr_{0.33}MnO_3$-$SrTiO_3$ superlattices are presented in the upper panel. Local magnetic moments of Ru and Mn are shown for both terminations in the lower panel. Dashed lines mark the corresponding bulk values.

### 3. X-ray Magnetic Circular Dichroism (XMCD)

X-ray absorption spectra at Mn and Ru L-edges were measured using circular polarized light with the photon spin parallel ($\sigma^+$) or antiparallel ($\sigma^-$) with respect to the magnetic field.[4] The spectra were collected with the beam in grazing ($H$ // film surface) and in normal ($H \perp$ film surface) incidence. Fig.S3 (top) shows the geometrical arrangement. The degree of circular polarization delivered by the Apple II-type elliptical undulator was 70 % for the Ru L-edge and close to 100 % for the Mn L-edge. The spectra were recorded using the total electron yield method (by measuring the sample drain current) in a chamber with a vacuum base pressure of $2 \cdot 10^{-10}$ mbar. The XMCD data were measured at $T$ = 60 K after cooling the sample in a field of 3 T. Fig. S3 (bottom) presents an example of Mn and Ru spectra recorded in normal incidence ($H$ // [001]) for both terminations. (Spectra for grazing incidence are shown in Fig.2a of the main text.) As the foundation of the present results, data from both magnetic layers could be recorded reliably in either stacking sequence. The results underline the feasibility of such measurements for buried layers in magnetically coupled thin film structures.

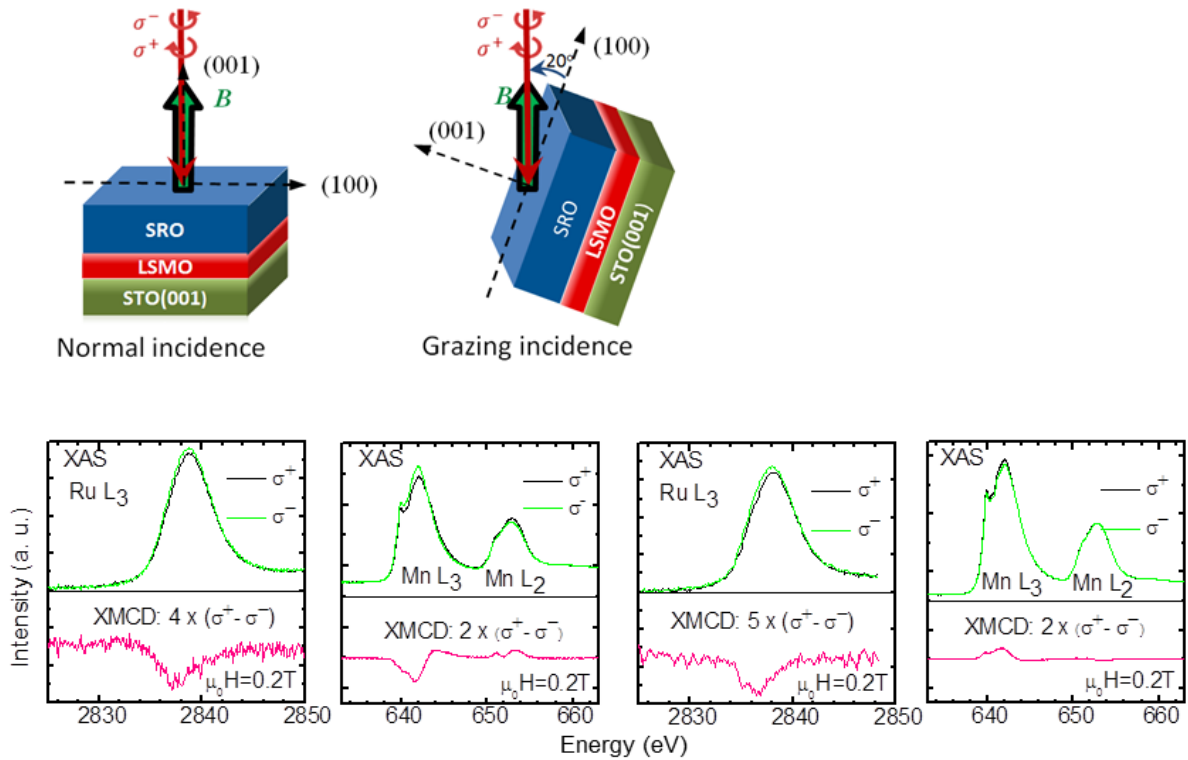

**Fig.S3: X-ray magnetic circular dichroism (XMCD) measurement geometry and representative out-of-plane spectra.** The magnetic field is applied parallel to the Poynting vector of the circularly polarized photons. Sample geometry for normal (H // [001]) and grazing (H // [100]) incidence, respectively, shown in the upper panel. Ru and Mn spectra recorded in normal incidence at 60 K and 0.2 T are shown in the lower panel (termination 1 (left) and termination 2 (right)).

The element-specific hysteresis loops for Mn and Ru measured at 60 K during applying a magnetic field *H* // [100] (in-plane, grazing incidence) or *H* // [001] (out-of-plane) are shown for termination 2 in Fig.S4 and for termination 1 in Fig.S5. The Ru-XMCD hysteresis loops were obtained by measuring the Ru $L_3$ edge XMCD spectrum for each value of magnetic field and integrating the XMCD signal. The Mn-XMCD hysteresis loops were obtained by measuring, as a function of applied field, the Mn $L_3$ edge XMCD signal at the energy where the XMCD signal is maximum. The direction of the magnetization in the layers is sketched in the inserted schemes at the loops in Fig.S4 and Fig.S5. (Note that the magnetization component perpendicular to the applied field has been chosen arbitrarily and kept constant in the schemes.) Termination 2 is discussed first, because it is the simpler case. The SRO layer in this type of sample is underneath the LSMO layer; this weakens the Ru signal and leads to some scattering of data as visible in Fig.S4. Nevertheless, Ru magnetic hysteresis has been recorded clearly. At 60 K, Ru moments reverse like in a typical ferromagnet with an out-of-plane canted magnetization orientation: in-plane (ip) and out-of-plane (oop) hysteresis loops have a large remanence. The coercive fields (+0.76 T / -0.62 T (oop) and +1.3 T / -1.2 T (ip)) reveal a small shift of the hysteresis curves to the positive field direction. Mn hysteresis loops are inverted for both field directions, indicating strong antiferromagnetic (afm) coupling to the SRO layer. (Positive orientation of Ru moments at remanence induces negative orientation of Mn moments.) Mn moments are canted out-of-plane, as is visible in the large remanence for both field directions in Mn hysteresis loops. Hence, the afm coupling to Ru induces canting of Mn moments, whereas LSMO in a single layer has pure in-plane magnetization. The inverted Mn hysteresis loops are also shifted slightly in positive field direction, reflected in their coercive fields (+0.45 T / -0.42 T (oop) and +0.12 T / -0.01 T (ip)). The loop shifts are likely to result from cooling the sample in a positive magnetic field of 3 T. Mn hysteresis loops show additional anomalies at the field values where Ru aligns with the applied field. At these anomalies, the Mn intensity drops as a consequence of the afm interaction of Mn and Ru which rotates Mn moments slightly away from the field direction.

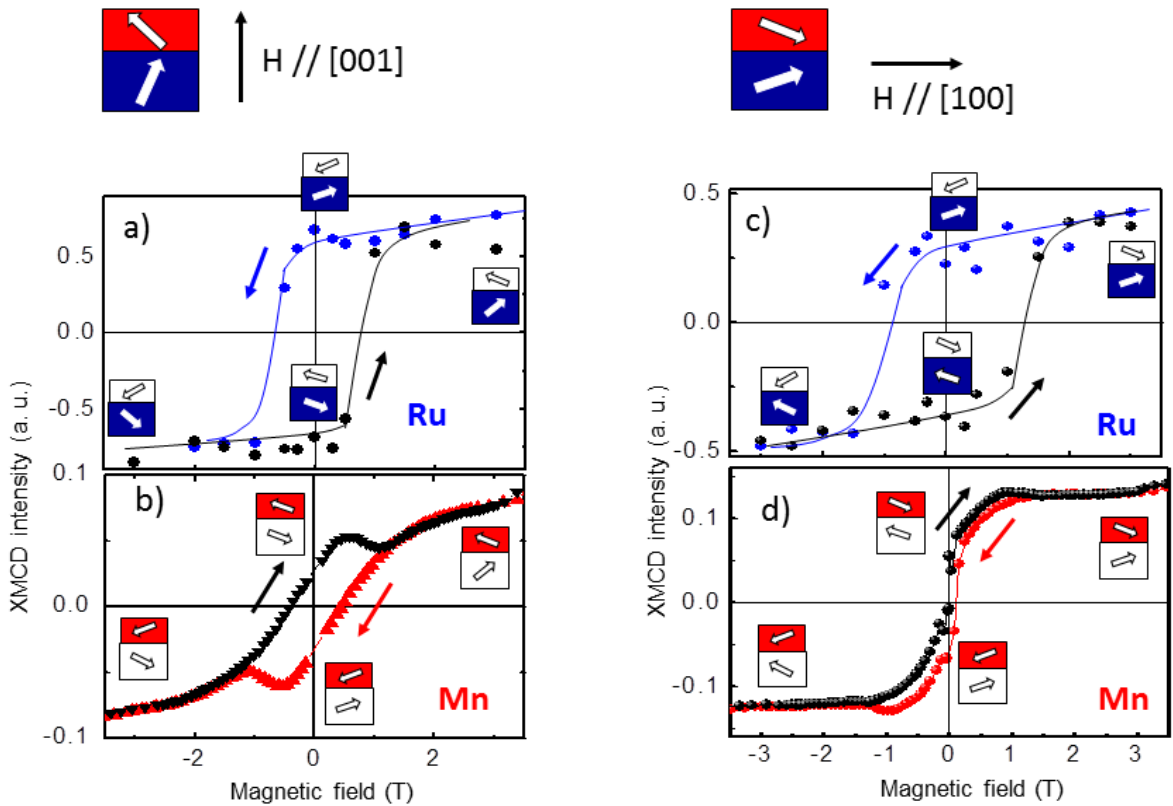

**Fig.S4: Element-specific hysteresis loops of Ru and Mn for termination 2 measured at 60 K.** Magnetic field applied in out-of-plane (H // [001]) (left side) and in-plane (H // [100]) direction (right side), respectively. In-plane data are identical with those in Fig.2 of the main text. Layer schemes around the hysteresis loop indicate the respective directions of the layer magnetizations. Lines in panels a,c) are guides to the eye.

Fig.S5 presents element-specific hysteresis loops for a sample of termination 1 type. Even though Mn spectra have been collected from a buried LSMO layer, they are distinct with very low scattering. As discussed in the main text, Mn moments show hysteresis characteristics similar to a single LSMO film. In the remanent state, Mn moments are oriented in-plane along [100] (or [010], for symmetry reasons). In-plane magnetization reversal occurs in low magnetic field (~30 mT, Fig.S5d). Out-of-plane magnetization arises from rotating Mn moments out of the film plane, requiring a field of 0.8-0.9 T for saturation (Fig.S5b). This behavior seems to indicate the absence of coupling between Mn and Ru moments at first glance. However, the Ru in-plane hysteresis loop (Fig.S5c) has a peculiar shape understood based on the presence of an interface-near layer which is antiferromagnetically aligned to the Mn moments in rigid way at small fields. This alignment can only be broken in larger magnetic field (of about 3 T at 140 K as derived from temperature-dependent magnetization measurements, see discussion of Fig.3 in the main text). The Ru in-plane hysteresis loop is discussed in detail in the main text. The Ru out-of-plane loop shows ferromagnetic hysteresis with large remanence, indicating substantial out-of-plane canting of Ru moments. This canting occurs in the upper part of the SRO layer, because interface-near Ru moments are in-plane oriented as is concluded from the absence of Mn canting. Interestingly, these results point to a peculiar non-collinear Ru order in the SRO layer at interfaces with termination 1, as suggested also in a neutron-depolarization study by Kim et al.[5] The non-collinear spin texture of Ru moments can be controlled by reorienting the Mn layer which is possible in small magnetic field.

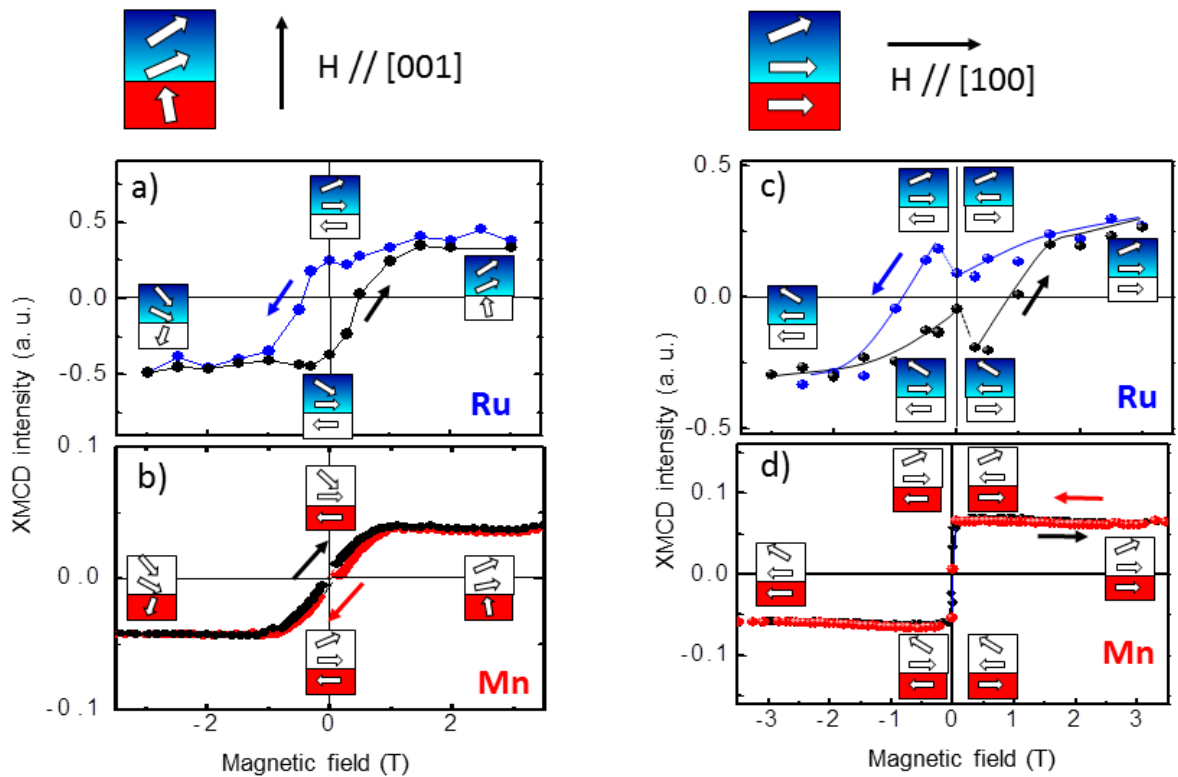

**Fig.S5: Element-specific hysteresis loops of Ru and Mn for termination 1 at 60 K.** Magnetic field applied in out-of-plane (H // [001]) (left side) and in-plane (H // [100]) direction (right side), respectively. In-plane data are identical with those in Fig.2 of the main text. Layer schemes around the hysteresis loop indicate the respective directions of the layer magnetizations. The Ru canting angle may change gradually between lower and upper part of the SRO layer. The line in panel c) is a guide to the eye.

*4. Magnetization measurements*

Fig.S6 shows representative magnetization loops taken at 10 K along [100] (in-plane) and [001] (out-of-plane) directions. Clear differences between termination 1 and 2 samples are visible which can be understood based on the XMCD results. We recall first that the true high-field slope of the magnetization cannot be derived from the present measurements, because the measured high-field slope d$M$/d$H$ of the samples contains a large diamagnetic substrate contribution. Therefore, it has been subtracted from the data. Further, the applied SQUID magnetometry tends to substantial error of the absolute value of the magnetic moment for measurements perpendicularly to the film plane. Hence, the saturation values obtained from the out-of-plane measurements should be seen with care. For termination 1, in-plane (ip) and out-of-plane (oop) hysteresis loops are slim with coercive fields of ~20 mT (ip) and 70 mT (oop). The slim in-plane loop underlines the coupled switching of Mn and Ru in a very small field as derived from XMCD data (Fig. 2 and Fig.S5). Ru-XMCD could not be measured in a very small field below 0.2 T (Fig. 2b), but the slim magnetization loop (Fig.S6, left panel) supports the scenario of rigidly coupled switching of Mn and Ru moments at the interface. There is some remanence in oop direction, in agreement with XMCD results, where the upper part of the SRO layer shows canted magnetization, but essentially, the oop hysteresis curve reflects a rotation process of magnetization which is finished at about 2 T. In in-plane data, there is a magnetization component that is aligned along the field in about 3 T. This component is attributed to aligning Ru moments to the field direction. However, the change observed in the magnetization at this transition is too small to account for a reversal of all Ru moments. If the interface-near layer can be reversed, i. e. antiferromagnetic Ru-Mn coupling can be broken in 4.5 T at 10 K, is not known at present. It is more likely that the upper part of the SRO layer aligns with the field. For termination 2, the oop hysteresis curve has large coercivity and remanence, in agreement with the canting of the entire SRO layer. The transition at about 2 T (oop) is likely to reflect the Ru reorientation which takes place at ~0.8 T at 60 K in XMCD data (Fig.S4a). The in-plane hysteresis loop (Fig.S6, right) shows nearly compensated magnetization at remanence where both layers are antiferromagnetically coupled. A close view into the small field range shows that the inner loop is inverted. According to XMCD results (Fig.S4c,d), Mn is aligned to the field first (below 1 T at 60 K), while Ru reverses in larger field. The Ru reversal is likely to cause the transition finished at 3.5 T in the curve *M*(*H*) taken at 10 K.

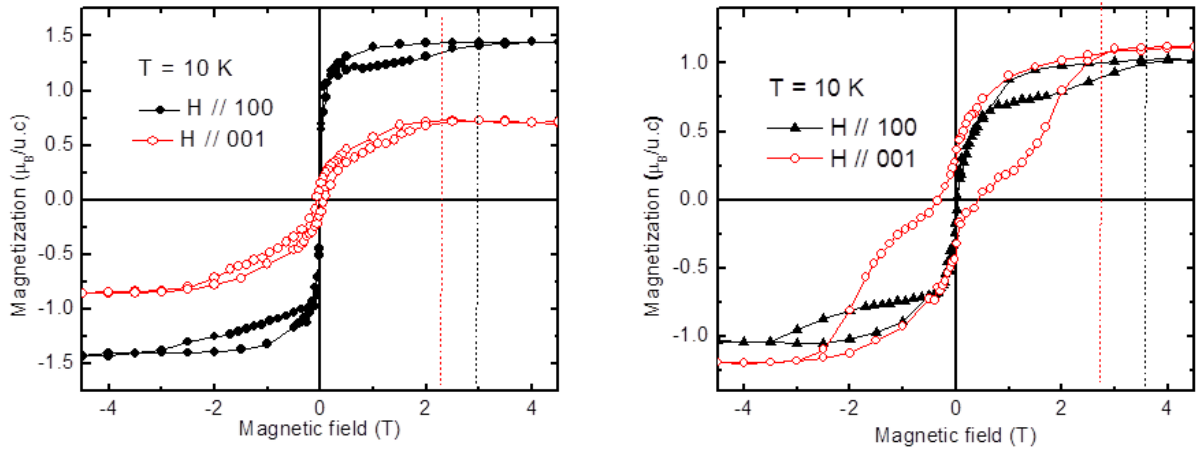

**Fig.S6: Field-dependent magnetization loops at 10 K for termination 1 (leftI) and termination 2 (right).** In-plane measurements with magnetic field $H$ // [100], out-of-plane measurements with $H$ // [001]. Dashed lines mark the fields where transitions are complete.

*Magnetization measurements of STO-capped samples*

Several pairs of bilayer samples with both terminations have been grown with an additional $SrTiO_3$ cap layer of 2 uc thickness. The goal was to probe samples which are close to the periodic LSMO-SRO-STO structure used for DFT calculations. Fig.S7 gives a representative example of the temperature-dependent magnetization curves of capped samples in analogy to Fig.3 in the main text. The magnetization data agrees well with that in Fig.3 in all essential features.

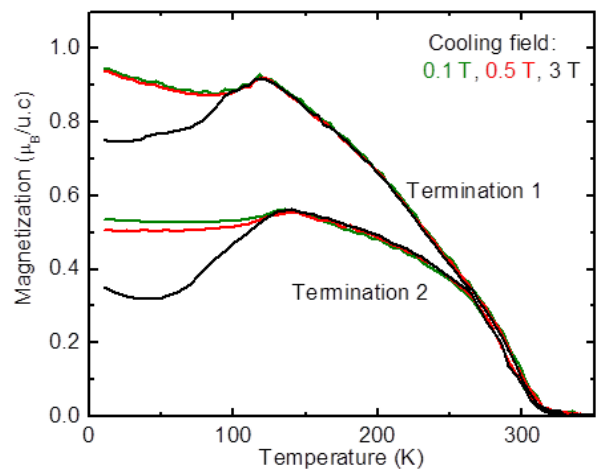

**Fig.S7: Temperature-dependent magnetization of samples with a $SrTiO_3$ cap layer of 2 unit cells.** Magnetization has been recorded in 0.1 T during warming, after the samples were cooled in the indicated fields.

## 5. Discussion of lattice symmetry and rotational patterns of oxygen octahedra at the LSMO-SRO interface

Collective rotations of oxygen octahedra ($MnO_6$ or $RuO_6$, with the transition metal in the centre of the octahedron) are crucially important for magnetic exchange interactions in oxides of $ABO_3$ perovskite type, because they are directly linked to the bond angles (Mn-O-Mn, Ru-O-Ru, Mn-O-Ru) and resulting orbital overlaps. In thin films, octahedral rotations are altered by the elastic strain state of the films [6-8] which is fixed in our bilayers through coherent growth on STO(001). Additionally, elastic interaction at an interface can induce cooperative rotations from one component layer into the next where the induced rotation decays within a penetration length of typically few unit cells.[9] The LSMO-SRO interface(s) have not yet been directly investigated regarding the octahedral rotation transfer. The rotational patterns in single LSMO/STO(001) and SRO/STO(001) films, however, are known and form a basis for discussion. LSMO with **a⁻a⁻a⁻** rotations (in Glazer notation) in the bulk phase was found to show suppressed rotations (**a⁰a⁰a⁰**) in a layer of 2-3 uc on cubic $TiO_2$-terminated STO(001) substrate, with subsequent change of the rotation pattern to **a⁺a⁻c⁰** for the next 10 uc.[8,10] The difference to the bulk rotation pattern is likely to arise from the tensile strain of LSMO (0.75%). Thus, the LSMO layer of 9 uc thickness in termination 1 samples is expected to have this rotational pattern. SRO with orthorhombic *Pbnm* symmetry has rotations of **a⁻b⁺c⁻** type (taking **b** as the long axis of the *Pbnm* unit cell). Coherently strained films keep this pattern for thicknesses of 5-80 nm on $TiO_2$-terminated STO(001) if the oxygen pressure during growth is not too low.[11] For the applied growth pressure, the SRO layer in termination 2 samples is thus expected to have the **a⁻b⁺c⁻** rotation type. The question about the nature of the interface-induced rotational pattern at the LSMO-SRO interface is open and quite fascinating: SRO with an in-plane lattice parameter of 3.905 Å is known to be close to a structural instability, changing lattice symmetry from the bulk-like orthorhombic (monoclinic) to tetragonal at an in-plane parameter of $a \sim 3.93$ Å.[12,13] The strain-induced tetragonal phase has [100] and [010] in-plane magnetic easy axes like the interface-near SRO layer in termination 1 samples. Hence, the observed magnetic anisotropy suggests tetragonal symmetry of SRO at that interface which would be associated with a different rotational pattern of $RuO_6$ octahedra. As argued in the main text, this symmetry change cannot be a result of the growth sequence of SRO on LSMO for termination 1, because SRO has tetragonal symmetry at the growth temperature of 700 °C [14] and undergoes the tetragonal-to-orthorhombic phase transition near 300 °C when it should not matter which of the interface partners has been grown first. (This conclusion is drawn under the assumption of negligible influence from the substrate which is fulfilled here, as discussed in the main text.) It is worth noting in this context that a structural change of SRO from orthorhombic to tetragonal was observed in coherent $Pr_{0.7}Ca_{0.3}MnO_3$/SRO superlattices on STO(001).[15]

*6. Point defects at the LSMO-SRO interface*

The most likely types of point defects at the LSMO-SRO interface are oxygen vacancies and interdiffused metal ions. Detection of oxygen vacancies at oxide interfaces is notoriously difficult, because appropriate methodological tools are yet in a pioneering stage.[16,17] Substantial oxygen vacancy enrichment in the SRO layers might be excluded, because Lu et al. showed that oxygen-deficient SRO films have a magnetic anisotropy with a magnetic easy axis perpendicular to the film plane.[11] This magnetic anisotropy has not been observed for the SRO layer in any of the studied samples. The modest Mn and Ru interdiffusion found by STEM at the termination 2 interface can affect the strength of the antiferromagnetic coupling. Therefore, further work should address the effect of a given level of interdiffusion on the coupling strength. However, since the DFT results predict the strongly different coupling for *interdiffusion-free* interfaces in agreement with experimental findings, we rule out the moderate interdiffusion found at the termination 2 interface as the origin of the weaker coupling.